\pdfoutput=1

\documentclass[11pt,twoside,a4paper,cmspaper,final,collab]{cms-tdr}
\def\svnVersion{62585:62643}\def\cmsCernNo{2011-063}\def\cmsCernDate{\today}\def\cmsMessage{Submitted to Physical Review D (Rapid Communications)}

\begin{document}\cmsNoteHeader{BPH-10-013}

\hyphenation{had-ron-i-za-tion}
\hyphenation{cal-or-i-me-ter}
\hyphenation{de-vices}
\RCS$Revision: 62643 $
\RCS$HeadURL: svn+ssh://alverson@svn.cern.ch/reps/tdr2/papers/BPH-10-013/branches/prd/BPH-10-013.tex $
\RCS$Id: BPH-10-013.tex 62643 2011-06-20 21:02:02Z alverson $

\def\invnb   {\ensuremath{\mbox{\,nb}^{-1}}}
\def\invpb   {\ensuremath{\mbox{\,pb}^{-1}}}
\def\nb   {\ensuremath{\,\mbox{nb}}}
\def\ub   {\ensuremath{\,\mu\mbox{b}}}
\def\um   {\ensuremath{\,\mu\mbox{m}}}
\def\tev  {\ensuremath{\mbox{\,TeV}}}
\def\gev  {\ensuremath{\mbox{\,GeV}}}
\def\gevc  {\ensuremath{\mbox{\,GeV}/c}}
\def\gevcc  {\ensuremath{\mbox{\,GeV}/c^2}}
\def\mev  {\ensuremath{\mbox{\,MeV}}}
\def\mevc  {\ensuremath{\mbox{\,MeV}/c}}
\def\mevcc  {\ensuremath{\mbox{\,MeV}/c^2}}
\def\tmu   {\ensuremath{\tt trackerMuon}}
\def\tmus   {\ensuremath{\tt trackerMuons}}
\def\cc   {\ensuremath{c\overline{c}}}
\def\pp {\ensuremath{\Pp\Pp}}
\def\BBs {\ensuremath{\cmsSymbolFace{B}_\cmsSymbolFace{s}^0}}
\def\BBd {\ensuremath{\PBz}}
\def\BBu {\ensuremath{\PBp}}
\def\BBBs {\ensuremath{\overline{\cmsSymbolFace{B}}_\cmsSymbolFace{s}^0}}
\def\psimumu {\ensuremath{\JPsi\to\Pgmp\Pgmm}}
\def\psik {\ensuremath{\PBp\to \JPsi \PKp}}
\def\psikst {\ensuremath{\PBz\to \JPsi \cmsSymbolFace{K}^{*0}}}
\def\psiphi {\ensuremath{\BBs\to \JPsi\ \phi}}
\def\phikk {\ensuremath{\phi \to \PKp \PKm}}
\def\MB {\ensuremath{M_{\PB}}}
\def\mB {\ensuremath{m_{\cPqb}}}
\def\dsdpt {\ensuremath{d\sigma/dp_\mathrm{T}^{\PB}}}
\def\dsdeta {\ensuremath{d\sigma/d\eta^{\PB}}}
\def\dsdy {\ensuremath{d\sigma/d y^{\PB}}}
\def\ptb {\ensuremath{p_\mathrm{T}^{\PB}}}
\def\pt {\ensuremath{p_\mathrm{T}}}
\def\pl {\ensuremath{p_\mathrm{L}}}
\def\etab {\ensuremath{\eta^{\PB}}}
\def\yb {\ensuremath{\left|y^{\PB}\right|}}
\def\ybb {\ensuremath{y^{\PB}}}
\newcommand\T{\rule{0pt}{2.6ex}}
\newcommand\B{\rule[-1.2ex]{0pt}{0pt}}

\cmsNoteHeader{BPH-10-013}
\title{\texorpdfstring{Measurement of the \BBs\ Production Cross Section with \psiphi\ Decays in
\pp\ Collisions at $\sqrt{s}$ = 7~\tev}{Measurement of the Strange B Meson Production Cross Section with J/Psi phi Decays in
pp Collisions at sqrt(s) = 7 TeV}}

\address[pu]{University of Tennessee}
\author[pu]{Giordano Cerizza}\author[pu]{Stefan Spanier}

\date{\today}

\abstract{
The \BBs\ differential production cross section is measured as functions of the transverse momentum and rapidity
in \pp\ collisions at $\sqrt{s} = 7\TeV$, using the \psiphi\ decay,
and compared with predictions based on perturbative QCD calculations at next-to-leading order.
The data sample, collected by the CMS experiment at the LHC, corresponds to an integrated
luminosity of $40\invpb$.
The \BBs\ is reconstructed from the decays \psimumu\ and $\phi\to \PKp \PKm$.
The integrated \BBs\ cross section times \psiphi\ branching fraction
in the range $8 < \ptb < 50$~GeV/$c$ and $\yb < 2.4$ is measured to be
$6.9\pm 0.6\pm 0.6$~nb, where the first uncertainty is statistical and
the second is systematic.
}

\hypersetup{
pdfauthor={CMS Collaboration},
pdftitle={Measurement of the Strange B Meson Production Cross Section with J/Psi phi Decays in
pp Collisions at sqrt(s) = 7 TeV},
pdfsubject={CMS},
pdfkeywords={CMS, B physics}}

\maketitle %maketitle comes after all the front information has been supplied

The measurements of differential cross sections for heavy-quark
production in high-energy hadronic interactions are critical input
for the underlying next-to-leading order (NLO) Quantum Chromodynamics
(QCD) calculations~\cite{NDE}.
While pro\-gress has been achieved in the understanding of heavy-quark
production at Tevatron energies~\cite{TeVI-CDF1,TeVI-CDF3,TeVI-CDF6,TeVI-D01,TeVI-D03,TeVI-D04,TeVII-CDF1,TEVII-CDF2,Cacciari04}, large theoretical uncertainties remain
due to the dependence on the renormalization and factorization scales.
Measurements of \cPqb-hadron production at the higher energies provided by
the LHC represent an important new test of theoretical approaches that aim
to reduce the scale dependence of NLO QCD calculations~\cite{Cacciari98,Kniehl08}.
The Compact Muon Solenoid (CMS) experiment, that covers a rapidity
range complementary to the specialised \cPqb-physics experiment
LHCb~\cite{CMSLHCb}, recently measured the cross sections
for production of \BBu~\cite{david} and \BBd~\cite{keith}
in \pp\ collisions at $\sqrt{s}=7\TeV$.
This paper presents the first measurement of the production of
\BBs, with \BBs\ decaying into $\JPsi\  \phi$,
and adds information to improve the understanding of \cPqb-quark production at this energy.
Data and theoretical predictions are compared at the level of NLO heavy-quark production.

The decay channel \psiphi\ is of wide interest for the
indirect search of physics beyond the standard model at the LHC. This decay proceeds
via the $\cPqb\to\cPqc\cPaqc\cPqs$ transition that probes the CP-violating phase
related to \BBs-\BBBs\ mixing. The standard model predicts this phase to be
close to zero~\cite{SM} while new phenomena may alter the observed phase~\cite{Dunietz}.
The differential production cross section analysis prototypes the signal extraction
towards time-dependent measurements in this final state.

A sample of exclusive \psiphi\ decays, with \psimumu\ and \phikk,
is reconstructed in the data collected in 2010 by the CMS
experiment, corresponding to an integrated luminosity of
$39.6\pm 1.6\invpb$. The differential production cross sections, $\dsdpt$ and $\dsdy$, are
determined as functions of the transverse momentum \ptb\ and rapidity \yb\ of the
reconstructed \BBs\ candidate.
Here, the rapidity \ybb\ is defined as $\frac{1}{2}\ln{\frac{E+c\pl}{E-c\pl}}$, where $E$
is the particle's energy and $\pl$ is the particle's momentum along the counterclockwise
beam direction. The differential cross sections are calculated from the measured signal
yields ($n_{\rm sig}$), corrected for
the overall efficiency ($\epsilon$), bin size ($\Delta x$, with $x =  \ptb,\yb$), and
integrated luminosity ($L$),
\begin{equation}
\frac{d\sigma(\Pp\Pp\to \psiphi)}{d x} = \frac{n_{\rm sig}}
{2\cdot \epsilon \cdot{\cal B}\cdot L\cdot\Delta x}\,\, ,
\label{eq:eq1}
\end{equation}
where ${\cal B}$ is the product of the branching fractions for the decays of the $\JPsi$ and $\phi$ mesons.
In each bin the signal yield is extracted with an unbinned maximum likelihood fit to the
$\JPsi\ \phi$ invariant mass
and proper decay length $ct$ of the \BBs\ candidates.
The factor of $2$ in Eq.~(\ref{eq:eq1}) accounts for the choice of quoting the cross section
for \BBs\ production,
while $n_{\rm sig}$ includes both \BBs\ and \BBBs\ states. The size of the bins is chosen such that the
statistical uncertainty on $n_{\rm sig}$ is comparable in each of them.

A detailed description of the CMS detector can be found elsewhere~\cite{JINST}.
The primary components used in this analysis are the silicon tracker and the muon
systems. The tracker operates in a $3.8$\,T axial magnetic field
generated by a superconducting solenoid having an internal diameter of $6$~m.
The tracker consists of three cylindrical layers of
pixel detectors complemented by two disks in the forward and backward
directions. The radial region between 20 and 116~cm is occupied
by several layers of silicon strip detectors in barrel and disk
configurations, ensuring at least nine hits in the pseudorapidity
range $|\eta| < 2.4$, where $\eta = -\ln{[\tan{(\theta/2)}]}$
and $\theta$ is the polar angle of the track relative to the
counterclockwise beam direction. An impact parameter resolution around
15\,$\mu$m and a \pt\ resolution around 1.5\,\% are achieved
for charged particles with transverse momenta up to $100$~GeV/$c$.
Muons are identified in the range $|\eta|< 2.4$, with detection planes made of
drift tubes, cathode strip chambers, and resistive plate chambers,
embedded in the steel return yoke.

The first level of the CMS trigger system
uses information from the crystal electromagnetic calorimeter, the
brass/scintillator hadron calorimeter, and the muon detectors to select the most
interesting events in less than $1~\mu$s. The high level trigger (HLT) employs software algorithms and
a farm of commercial processors to further decrease the event rate using information
from all detector subsystems. The events used in the measurement reported in this paper were collected with a
trigger requiring the presence of two muons at the HLT, with no explicit momentum
threshold.

Reconstruction of \psiphi\ candidates begins by identifying \psimumu\ decays.
The muon candidates must have one or more reconstructed segments in the muon system that match the extrapolated position of a track
reconstructed in the tracker. Furthermore, the muons are required to
lie within a kinematic acceptance region defined
as: $\pt^{\mu}> 3.3$~GeV/$c$ for $|\eta^{\mu}| < 1.3$;
total momentum $p^{\mu}> 2.9$~GeV/$c$ for $1.3 < |\eta^{\mu}| < 2.2$;
and $\pt^{\mu}> 0.8$~GeV/$c$ for $2.2 < |\eta^{\mu}| < 2.4$.
Two oppositely charged muon candidates are paired and are required to originate from a
common vertex using a Kalman vertex fit.
The muon pair is required to have a transverse momentum $\pt > 0.5$~GeV/$c$
and an invariant mass within 150~MeV/$c^2$ of the world average $\JPsi$ mass value~\cite{PDG2010},
which corresponds to more than three times the measured dimuon invariant mass resolution~\cite{BPH-10-002}.

Candidate $\phi$ mesons are reconstructed from pairs of oppositely
charged tracks with $\pt>0.7$~GeV/$c$ that are selected from a
sample with the muon candidate tracks removed.
The tracks are required to have at least five hits in the silicon tracker detectors,
and a track $\chi^2$ per degree of freedom less than five.
Each track is assumed to be a kaon and the invariant mass
of a track pair has to be within $10$~MeV/$c^2$
of the world average $\phi$-meson mass~\cite{PDG2010}.

The \BBs\ candidates are formed by combining a $\JPsi$ with a $\phi$ candidate.
The two muons and the two kaons are subjected to a combined vertex and kinematic fit~\cite{kinfit}, where in addition
the dimuon invariant mass is constrained to the nominal $\JPsi$ mass.
The selected candidates must have a resulting $\chi^2$ vertex probability
greater than $2\%$, an invariant mass between $5.20$ and $5.65$~GeV/$c^2$,
and be in the kinematic range $8< \ptb <50$~GeV/$c$ and $\yb<2.4$.
For events with more than one candidate, the one with the highest
vertex-fit probability is selected, which results in the correct
choice 97\% of the time, as determined from simulated signal events.

The proper decay length of each selected \BBs\ candidate
is calculated using the formula $ct = c(\MB/ \ptb)L_{xy}$, where the transverse decay
length $L_{xy}$ is the length of the vector $\vec{s}$ pointing from the primary
vertex~\cite{TRK-10-005} to the secondary vertex projected onto the
\BBs\ transverse momentum:
$L_{xy} = (\vec{s}\cdot\vec{p}_T^{\PB})/{\ptb}$, with $\MB$ being the reconstructed mass
of the \BBs\ candidate.
Candidate \BBs\ mesons are accepted within the range $-0.05 < ct < 0.35$~cm.

A total of $6200$ events pass all the selection criteria.
The efficiency of the \BBs\ reconstruction is computed with a combination of techniques using the data and
large samples of simulated signal events generated using \PYTHIA\ $6.422$~\cite{PYTHIA}.
The decays of unstable particles are described by the \EVTGEN~\cite{EvtGen} simulation.
Long-lived particles are then propagated through a detailed description of the CMS detector based on the \GEANTfour~\cite{GEANT4}
package. The trigger and muon-reconstruction efficiencies are obtained from a large sample of inclusive
$\psimumu$ decays in data using a (tag-and-probe) technique similar to that described in Ref.~\cite{BPH-10-002},
where one muon (the tag) is identified with stringent quality requirements,
and the second muon (the probe) is identified using information either exclusively from the
tracker (to measure the trigger and muon-identification efficiencies), or from the
muon system (to measure the silicon tracking efficiency).
The dimuon efficiencies are calculated as the product of the single-muon efficiencies obtained with this method.
Corrections to account for correlations between the two muons (1--3\%) are obtained from simulation studies.
The correction factors are determined in bins of single muon $\pt^{\mu}$ and $\eta^{\mu}$, and are applied
independently to each muon from a \psiphi\ decay in the simulation to determine
the total corrected efficiency.
The probabilities for the muons to lie within the kinematic acceptance region and for the
$\phi$ and \BBs\ candidates to pass the selection requirements are determined from the simulated events.
The efficiencies for hadronic track reconstruction~\cite{TRK-10-002} and the vertex-quality requirement are found to
be consistent between real data and simulated events within their uncertainties (up to $5\%$).
The total efficiency of this selection, defined as the fraction of \psiphi\
decays produced with $8< \ptb <50$~GeV/$c$ and $\yb<2.4$ that pass all criteria, ranges from
1.3\% for $\ptb\approx\,8$~GeV/$c$ to $19.6\%$ for $\ptb>23$~GeV/$c$.

Backgrounds are dominated by prompt and non-prompt $\JPsi$ production.
The latter background is dominated by \PBp\ and \PBz\ mesons that decay to a
\JPsi\ and a higher-mass K-meson state (such as the $\PKp_1$). Such events
tend to contribute to the low-mass side of the \MB\ mass distribution.
Inspection of a large variety of potential background channels confirms that
there is no single dominant component and that the channel
$\PBz\to \JPsi\ \PKst$ (with $\PKst^0\to \PKp\pi^-$),
which \textit{a priori} is kinematically similar to the signal decay and more
abundantly produced, is strongly suppressed
by the restriction on the $\PKp\PKm$ invariant mass.
A study of the sidebands of the dimuon invariant mass distribution confirms that
the contamination from events without a $\JPsi$ decay to two muons is negligible after all selection
criteria have been applied.

The signal yields in each \ptb\ and \yb\ bin, given in Table~\ref{tab:results},
are obtained using an unbinned extended maximum-likelihood fit to $\MB$ and $ct$.
The likelihood for event $j$ is obtained by summing the product of the yield $n_i$ and
the probability density functions (PDF) ${\cal P}_i$ and ${\cal Q}_i$ for each of the signal and background hypotheses
$i$.  Three individual components are considered: signal,
non-prompt b $\to\JPsi$ X, and prompt $\JPsi$.  The extended likelihood function
is then the product of likelihoods for each event $j$:
\begin{equation}
{\cal L} = \exp\left ( -\sum_{i=1}^{3} n_i \right ) \prod_j \left [
\sum_{i=1}^{3} n_i{\cal P}_i(\MB;\, \vec{\alpha}_i){\cal Q}_i(ct;\, \vec{\beta}_i) \right ].
\end{equation}
The PDFs ${\cal P}_i$ and ${\cal Q}_i$ are parameterized separately for each fit component
with shape parameters $\vec{\alpha}_i$ for $\MB$ and $\vec{\beta}_i$ for $ct$.
The yields $n_i$ are then determined by minimizing the quantity $-\ln{\cal L}$ with
respect to the signal yields and a subset of the PDF parameters~\cite{roofit}.
Possible correlations between $\MB$ and $ct$ are found to be
less than $2\%$.  Therefore, they are assumed to have a negligible impact
on the fit, and potential biases arising from this assumption are accounted for in the
systematic uncertainty on the fitted signal yield as described below.

\begin{table*}[htbp]
\caption{ Signal yield $n_{\rm sig}$, efficiency $\epsilon\,(\%)$, and measured differential cross sections $\dsdpt$
and $\dsdy$, compared to the \MCATNLO and \PYTHIA\ predictions, in different \ptb\ and \yb\ intervals.
The uncertainties on the measured cross sections are statistical and systematic, respectively, excluding
the common luminosity of $4\%$ and the $1.4\%$ from the $\JPsi$ and $\phi$ branching fractions.}
\label{tab:results}
\begin{center}
\setlength{\extrarowheight}{2.5pt}
\begin{tabular}{c@{\hspace{2.6ex}}r@{$\ \pm\ $}lr@{$\ \pm\ $}lccc} \hline
$\ptb$ (\gevc)  & \multicolumn{2}{c}{$n_\text{sig}$} & \multicolumn{2}{c}{$\epsilon$ (\%)} & \multicolumn{3}{c}{$\dsdpt$ $($nb$/\gevc)$}\\
 & \multicolumn{2}{c}{} & \multicolumn{2}{c}{} & Data & \MCATNLO & \PYTHIA\ \\ \hline

$8$--$12$   & 138&16  & 1.28&0.05   & $1.172  \pm 0.136 \pm 0.113$   & $0.719$   & $1.513$ \\
$12$--$16$  & 176&17  & 5.26&0.23   & $0.364  \pm 0.035 \pm 0.034$   & $0.240$   & $0.515$ \\
$16$--$23$  & 162&16  & 11.9&0.6    & $0.085 \pm 0.008 \pm 0.008$    & $0.074$   & $0.144$ \\
$23$--$50$  & 86&11   & 19.6&1.1    & $0.007 \pm 0.001 \pm 0.001$    & $0.008$   & $0.010$ \\ \hline
$\yb$ & \multicolumn{2}{c}{$n_\text{sig}$} & \multicolumn{2}{c}{$\epsilon$ (\%)} & \multicolumn{3}{c}{$\dsdy$  $($nb$)$}\\
   & \multicolumn{2}{c}{} & \multicolumn{2}{c}{} & Data & \MCATNLO & \PYTHIA\ \\\hline
$0.00$--$0.80$  & 151&15  & 2.75&0.09 & $1.484 \pm 0.147 \pm 0.148$    & $1.040$    & $2.281$  \\
$0.80$--$1.40$  & 144&15  & 4.65&0.18 & $1.123 \pm 0.117 \pm 0.102$    & $1.023$    & $2.051$  \\
$1.40$--$1.70$  & 129&15  & 5.68&0.31 & $1.634 \pm 0.190 \pm 0.160$    & $0.929$    & $1.833$  \\
$1.70$--$2.40$  & 139&17  & 3.26&0.20 & $1.316 \pm 0.161 \pm 0.139$    & $0.801$     & $1.559$  \\ \hline
\end{tabular}
\end{center}
\end{table*}

The PDFs are constructed from basic analytical functions that satisfactorily describe
the variable distributions from simulated events. Shape parameters are
obtained from data when possible. The $\MB$ PDF is the sum of two
Gaussian functions for the signal, a second-order polynomial for the non-prompt
$\JPsi$ that allows for possible curvature in the shape, and a first-order
polynomial for prompt $\JPsi$. The resolution on $\MB$ is approximately $20$~MeV/$c^2$ near the \BBs\ mass.

For the signal, the $ct$ PDF is a single exponential parameterized in terms of
a proper decay length $c\tau$. It is convolved with a resolution
function that is a combination of two Gaussian functions to account for
a dominant core and small outlier distribution; the core fraction is
varied in the fit and found to be consistently larger than 95\%.
The $ct$ distribution for the non-prompt $\JPsi$ background is described
by a sum of two exponentials, with effective lifetimes that are allowed
to be different.
The ``long-lifetime exponential'' corresponds to decays of \cPqb-hadrons to
a \JPsi\ plus some charged particles that survive the $\phi$ selection, while the
``short-lifetime exponential'' accounts for events where the muons from the
\JPsi\ decay are wrongly combined with hadron tracks originating from the
\pp\ collision point.
The exponential functions are convolved with a resolution function
with the same parameters as the signal.
For the prompt $\JPsi$ component the pure resolution function is used.
The core resolution in $ct$ is measured in data to be $45\ \mu$m.

All background shapes are obtained directly from data, while the signal shape in $\MB$ is taken from
a fit to reconstructed signal events from the simulation.
The effective lifetime and resolution function parameters for prompt and non-prompt
backgrounds are extracted, using the full data sample irrespectively of \ptb\ and \yb,
from regions in $\MB$ that are separated
by more than four times the width of the observed \BBs\ signal from the mean
\BBs\ peak position ($\MB$ sidebands): $5.20 < \MB\ < 5.29$~GeV/$c^2$ and
$5.45 < \MB\ < 5.65$~GeV/$c^2$. A comparison of the PDF shapes for the different
sideband regions in simulated events confirms that their average over the signal-free
regions is a good representation of the background in the signal region.
With the lifetimes for signal and non-prompt background fixed from this first step, the
resolution function parameters are then determined separately in each
\ptb\ and \yb\ bin, from the $\MB$ sidebands.
The signal and background yields in each \ptb\ and \yb\  bin are determined
in a final iteration, using the full \MB\ range, with all parameters floating
except the background lifetimes and the lifetime resolution functions, which
are fixed to the results of the fit to the \MB\ sidebands.
It has been verified that leaving all parameters floating
changes the signal yield by an amount smaller than the systematic
uncertainty assigned to the fit procedure.

Many detailed studies have been conducted to validate the accuracy and
robustness of the fit procedure. A large number of pseudo-experiments were performed, each
corresponding to the yields observed in each \ptb\ and \yb\ bin for
a data sample corresponding to an integrated luminosity of $40\invpb$, where signal and background
events were generated randomly from the PDFs in each bin. The fit yields were found to be unbiased
and their uncertainties estimated properly. The effects of residual correlations between $\MB$ and $ct$
were studied by mixing fully simulated signal and background events to produce pseudo-experiments.
The observed deviations between the fitted and generated yields (1--2\%) are taken as the
systematic uncertainty due to potential biases in the fit method.

\begin{figure}[htbp]
  \begin{center}
   \includegraphics[width=0.95\columnwidth]{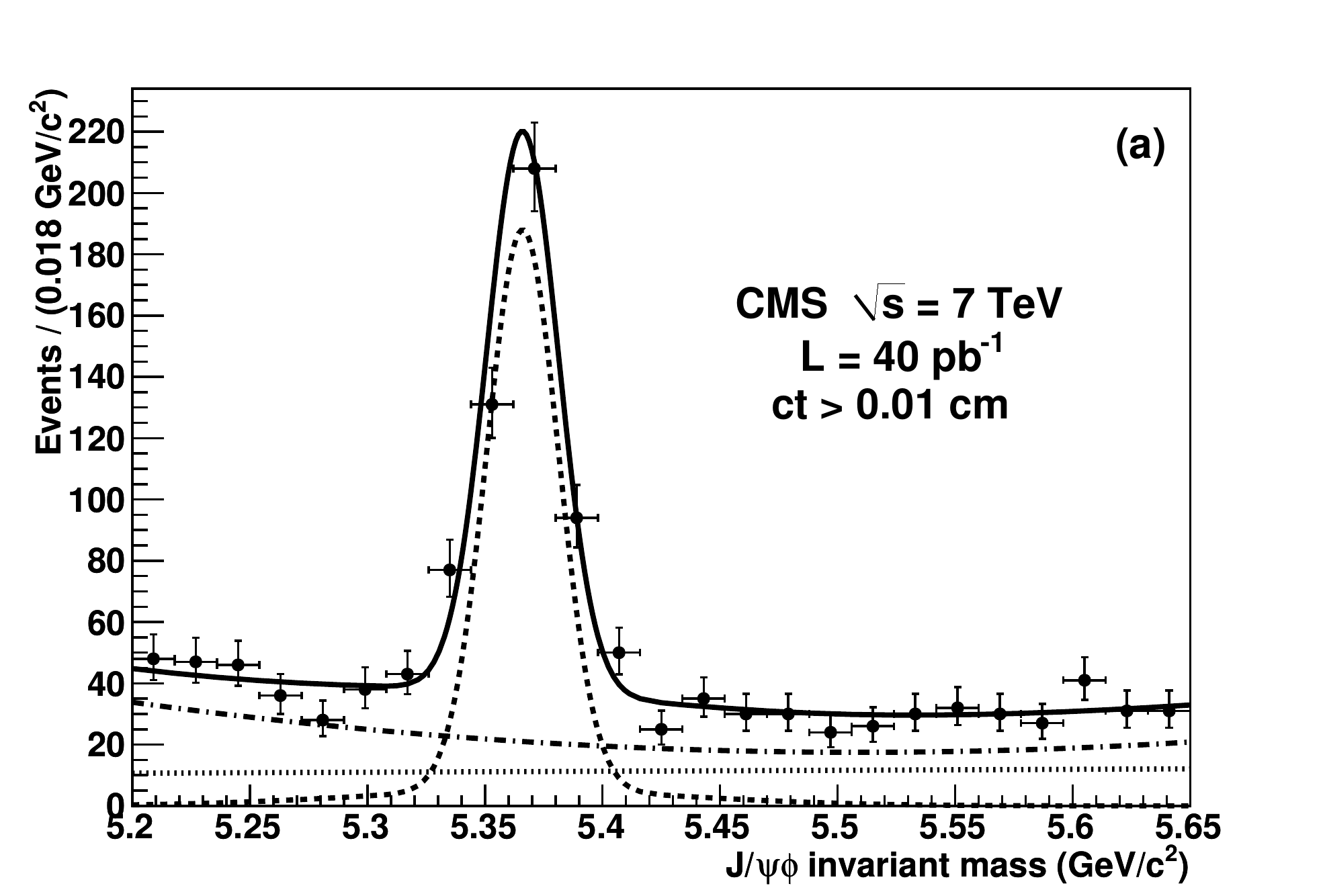}
   \includegraphics[width=0.95\columnwidth]{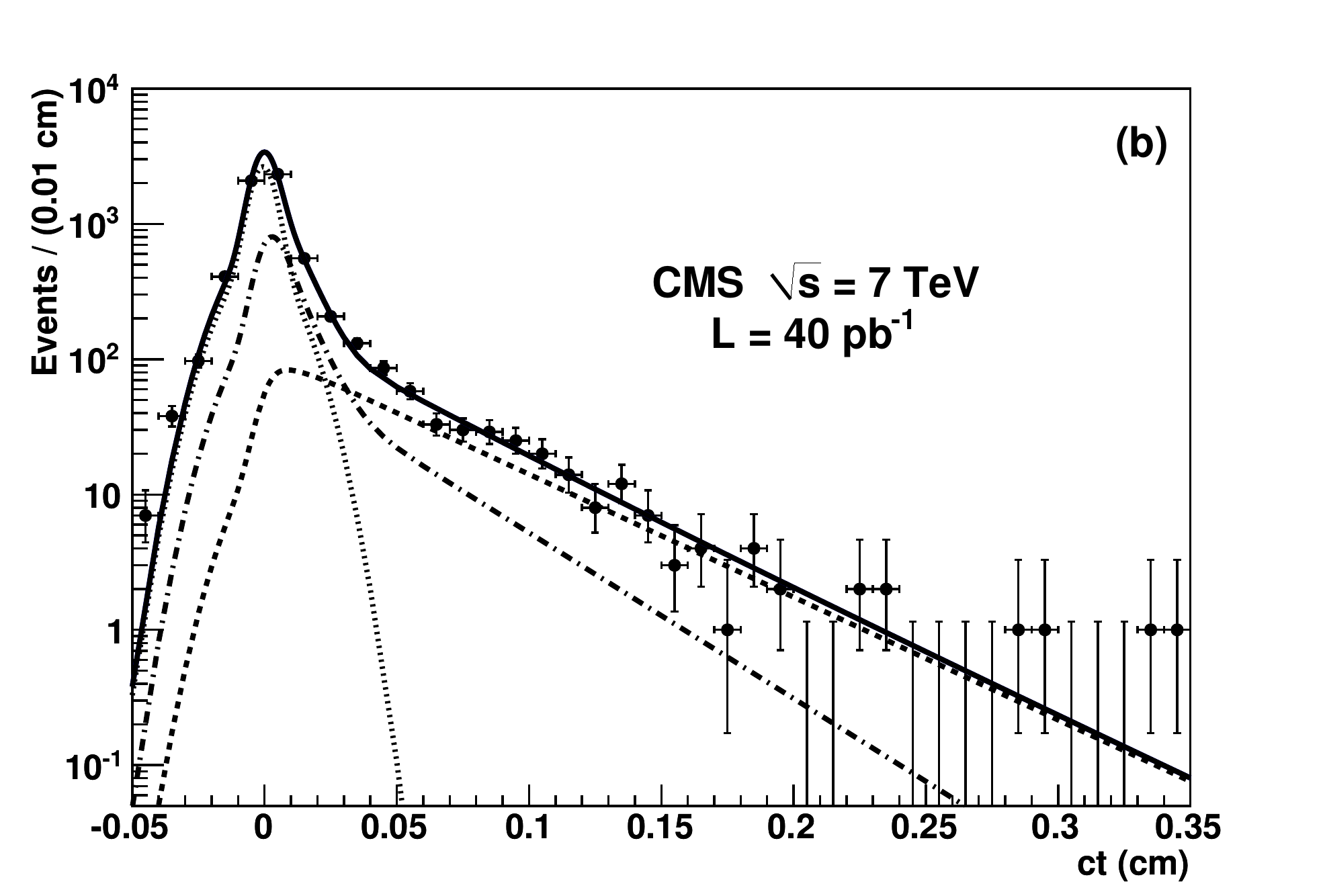}
    \caption{Projections of the fit results in $\MB$ (a) and $ct$ (b) for
    $8< \ptb <50$~GeV/$c$ and $\yb < 2.4$. The curves in each plot are: the sum of all
    contributions (solid line); signal (dashed); prompt $\JPsi$ (dotted);
    and non-prompt $\JPsi$ (dot-dashed).
    For better visibility of the individual contributions, plot (a) includes the requirement $ct>0.01\ {\rm cm}$.}
   \label{fig:fit}
  \end{center}
\end{figure}

Figure~\ref{fig:fit} shows the fit projections for $\MB$ and $ct$ from the
inclusive sample with $8< \ptb <50$~GeV/$c$ and $\yb < 2.4$.  When plotting $\MB$,
the selection $ct>0.01$ cm is applied for better visibility of the individual
contributions. %, which are scaled by the efficiency of this additional requirement.
The number of signal events in the entire data sample is $549\pm 32$, where the uncertainty is
statistical only. The obtained proper decay length of the signal, $c\tau = 478 \pm 26$ $\mu$m,
is within $1.4$ standard deviations of the world average value~\cite{PDG2010}, even though
this analysis was not optimized for lifetime measurements.

Table~\ref{tab:results} summarizes the fitted signal yield in each
bin of \ptb\ and \yb. The differential cross section is calculated according to
Eq.~(\ref{eq:eq1}), using the product of the branching fractions ${\cal B}(\psimumu)=(5.93\pm 0.06)\times 10^{-2}$
and ${\cal B}(\phikk)=(48.9\pm 0.5)\times 10^{-2}$~\cite{PDG2010}.
All efficiencies are calculated separately in each bin, and account for bin-to-bin
migrations (less than $1\%$) due to the finite resolution of the measured momentum and rapidity.

\begin{figure}[htbp]
\begin{center}
\includegraphics[width=0.95\columnwidth]{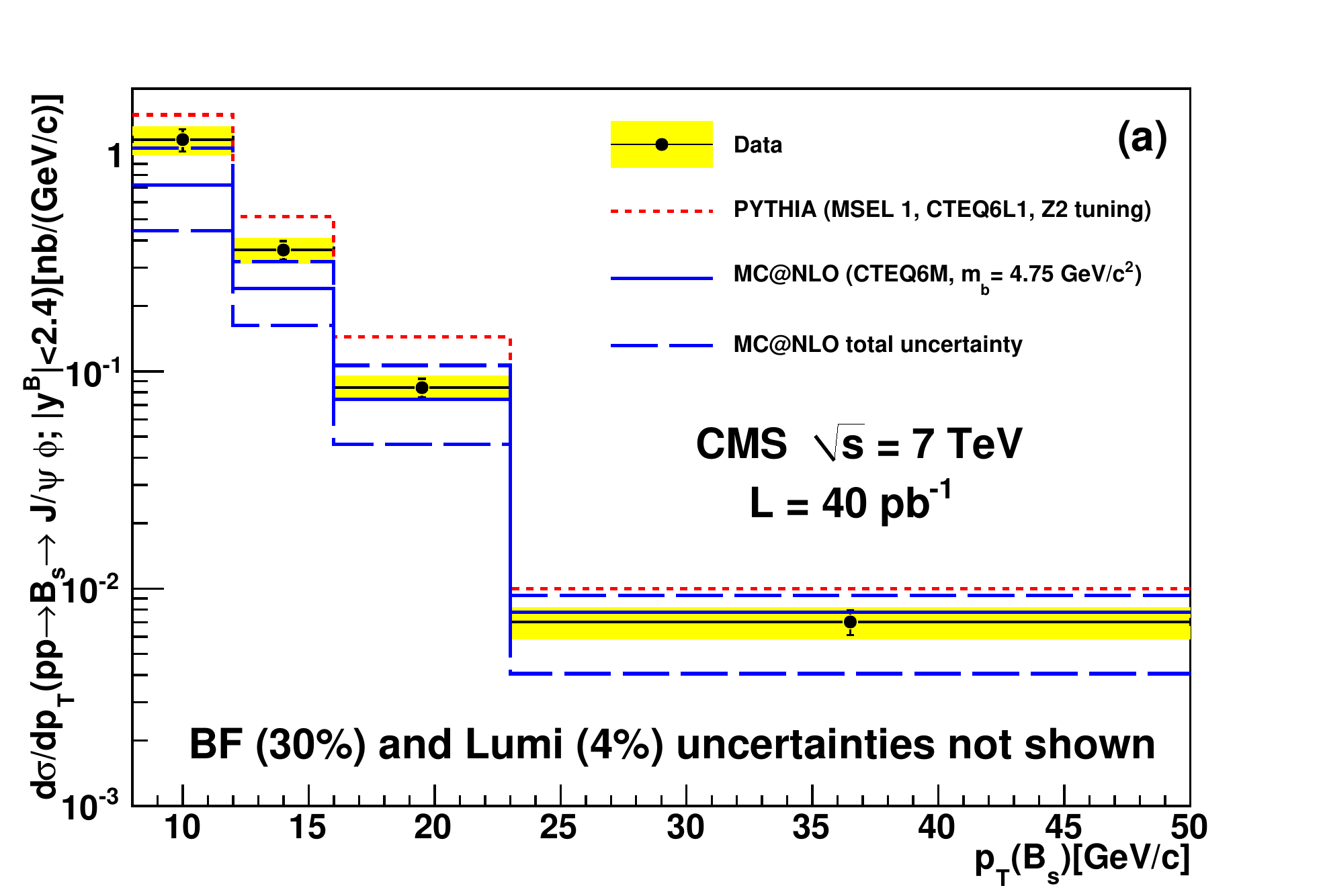}
\includegraphics[width=0.95\columnwidth]{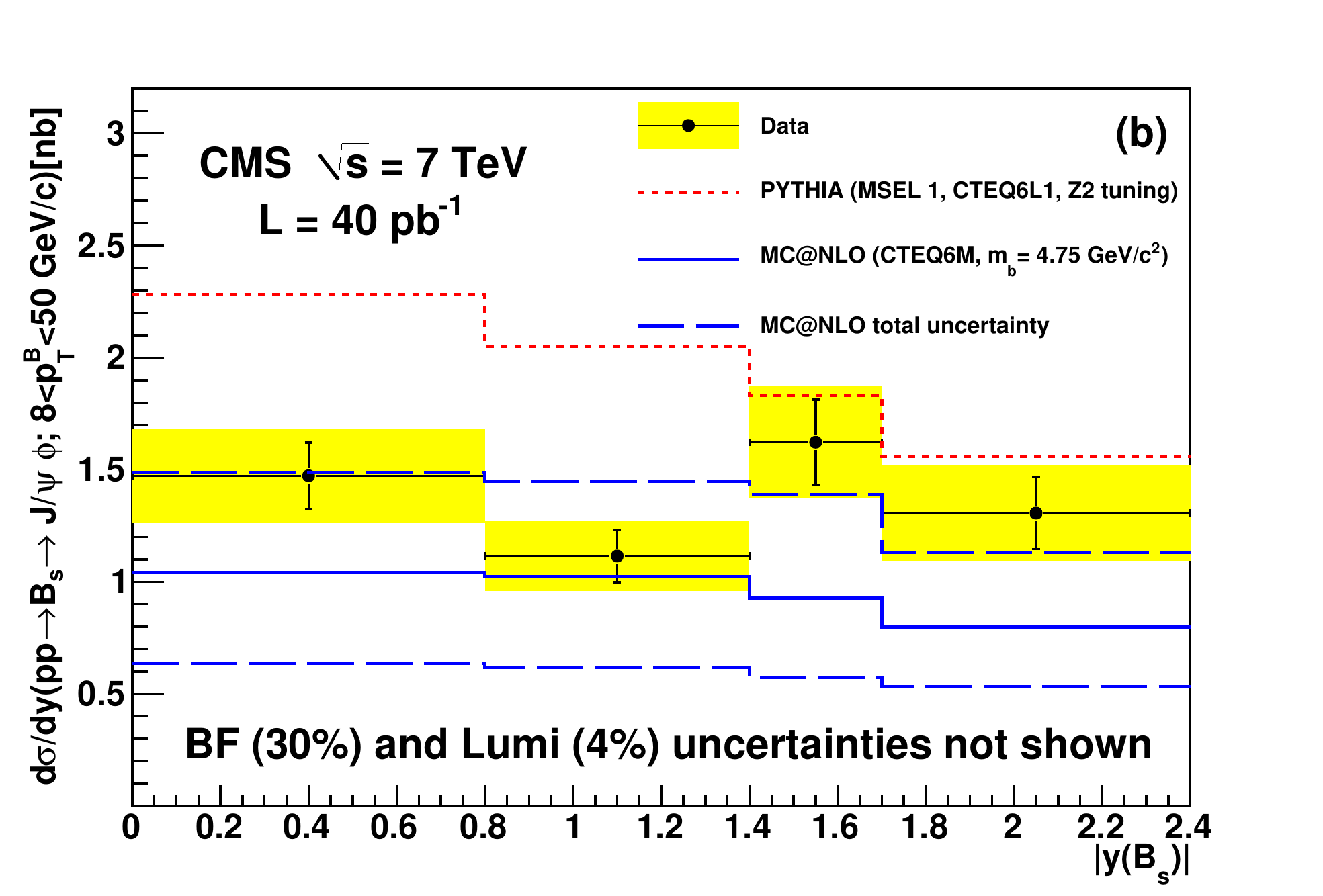}
\caption{Measured differential cross sections $\dsdpt$ (a) and $\dsdy$ (b) compared with
theoretical predictions.  The (yellow) band represents the sum in
quadrature of statistical and systematic uncertainties.
The dotted (red) line is the \PYTHIA\ prediction; the solid and
dashed (blue) lines are the \MCATNLO prediction and its uncertainty, respectively.
The common uncertainties of $4\%$ on the data points, due to the integrated luminosity,
and of $30\%$ on the theory curves, due to the \psiphi\ branching fraction, are not shown.}
\label{fig:Xsec}
\end{center}
\end{figure}

The cross section measurement is affected by several sources of systematic uncertainty arising
from uncertainties on the fit, efficiencies, branching fractions,
and integrated luminosity. In every bin the total uncertainty is about $11\%$.
Uncertainties on the muon efficiencies from the trigger, identification, and tracking
are determined directly from data (3--5\%). The uncertainty of the method
employed to measure the efficiency in the data has been
estimated from a large sample of full-detector simulated events (1--3\%).
The tracking efficiency for the charged kaons has been shown to be
consistent with simulation. A conservative uncertainty of at most $9\%$
in each bin has been assigned for the hadronic track reconstruction (adding linearly
the uncertainties on the two kaon tracks~\cite{TRK-10-002}), which includes the
uncertainty due to misalignment of the silicon detectors.
The uncertainty on the fit procedure arising from potential
biases and imperfect knowledge of the PDF parameters is estimated by varying the parameters
by one standard deviation (2--4\%).
The contribution related to the \BBs\ momentum spectrum (1--3\%) is evaluated
by reweighting the shape of the \ptb\ distribution generated with \PYTHIA\ to match
the spectrum predicted by \MCATNLO~\cite{MCNLO}. An uncertainty of $1\%$ is assigned to the
variation of the selection criteria applied to the vertex fit probability, the
transverse momentum of the kaons, the \BBs\ transverse momentum, and the $\PKp\PKm$ invariant mass window.
An uncertainty is added to account for the limited number of simulated events
(at most $3\%$ in the highest \ptb\ bin).
The total uncorrelated systematic uncertainty on the cross-section measurement is computed in each bin
as the sum in quadrature of the individual uncertainties, and is summarized in Table~\ref{tab:results}.
In addition, there are common uncertainties of $4\%$ from the integrated luminosity measurement~\cite{EWK-10-004} and
$1.4\%$ from the $\JPsi$ and $\phi$ branching fractions.
As the reported result is a measurement of the \BBs\ cross section times the
\psiphi\ branching fraction,
the $30\%$ uncertainty on the
\psiphi\ branching fraction~\cite{PDG2010} is not included in the result.

The differential cross sections times branching fraction as functions of \ptb\ and \yb\ are listed in Table~\ref{tab:results}
and plotted in Fig.~\ref{fig:Xsec}, together with predictions from \MCATNLO and   \PYTHIA .
The predictions of \MCATNLO\  use the renormalization and factorization scales $\mu=\sqrt{\mB^2c^4+\pt^2c^2}$, where \pt\ is
the transverse momentum of the \cPqb\ quark, a \cPqb-quark mass of $\mB = 4.75$~GeV/$c^2$, and the CTEQ6M parton distribution
functions~\cite{CTEQ}. The uncertainty on the \MCATNLO\ cross section is obtained simultaneously varying the renormalization and
factorization scales by factors of two, varying \mB\ by $\pm 0.25$~GeV/$c^2$, and using the CTEQ6.6 parton distribution
function set. The prediction of \PYTHIA\ uses the CTEQ6L1 parton distribution
functions~\cite{CTEQ}, a \cPqb-quark mass of $4.8$~GeV/$c^2$, and the Z2 tune~\cite{Z2} to
simulate the underlying event.
The total integrated \BBs\ cross section times \psiphi\ branching fraction for the
range $8 < \ptb < 50$~GeV/$c$ and $\yb < 2.4$ is
measured to be $6.9\pm 0.6\pm 0.6$~nb, where the first uncertainty is statistical and the second is systematic.
The statistical and systematic uncertainties are derived from the bin-by-bin uncertainties and propagated through the sum. The result lies between the theoretical predictions of \MCATNLO\ ($4.6^{+1.9}_{-1.7}\pm 1.4$~nb)
and \PYTHIA\ ($9.4\pm2.8$~nb), where
the last uncertainty is from the \psiphi\ branching fraction~\cite{PDG2010}.
Also the previous CMS cross-section measurements
of $\PBp$~\cite{david} and $\PBz$~\cite{keith} production in \pp\ collisions at $\sqrt{s}=7\TeV$,
gave values between the two theory predictions, indicating internal consistency amongst the three
different \PB-meson results.

In summary, the first measurements of the \BBs\ differential cross sections $\dsdpt$ and $\dsdy$,
in the decay channel \psiphi\
and in $\Pp\Pp$ collisions at $\sqrt{s}=7\TeV$, have
been presented.  The results cover the kinematical window $\yb < 2.4$ and
$8<\ptb\ <50$~GeV/$c$. They add complementary information to previous results in moving towards
a comprehensive description of \cPqb-hadron production at $\sqrt{s}=7\TeV$.

We wish to congratulate our colleagues in the CERN accelerator departments for the
excellent performance of the LHC machine. We thank the technical and administrative
staff at CERN and other CMS institutes, and acknowledge support from: FMSR (Austria);
FNRS and FWO (Belgium); CNPq, CAPES, FAPERJ, and FAPESP (Brazil); MES (Bulgaria); CERN;
CAS, MoST, and NSFC (China); COLCIENCIAS (Colombia); MSES (Croatia); RPF (Cyprus);
Academy of Sciences and NICPB (Estonia); Academy of Finland, ME, and HIP (Finland);
CEA and CNRS/IN2P3 (France); BMBF, DFG, and HGF (Germany); GSRT (Greece);
OTKA and NKTH (Hungary); DAE and DST (India); IPM (Iran); SFI (Ireland); INFN (Italy);
NRF and WCU (Korea); LAS (Lithuania); CINVESTAV, CONACYT, SEP, and UASLP-FAI (Mexico);
PAEC (Pakistan); SCSR (Poland); FCT (Portugal); JINR (Armenia, Belarus, Georgia,
Ukraine, Uzbekistan); MST and MAE (Russia); MSTD (Serbia); MICINN and CPAN (Spain);
Swiss Funding Agencies (Switzerland); NSC (Taipei); TUBITAK and TAEK (Turkey);
STFC (United Kingdom); DOE and NSF (USA).

\bibliography{auto_generated}	% will be created by the tdr script.
\cleardoublepage\appendix\section{The CMS Collaboration \label{app:collab}}\begin{sloppypar}\hyphenpenalty=5000\widowpenalty=500\clubpenalty=5000\textbf{Yerevan Physics Institute,  Yerevan,  Armenia}\\*[0pt]
S.~Chatrchyan, V.~Khachatryan, A.M.~Sirunyan, A.~Tumasyan
\vskip\cmsinstskip
\textbf{Institut f\"{u}r Hochenergiephysik der OeAW,  Wien,  Austria}\\*[0pt]
W.~Adam, T.~Bergauer, M.~Dragicevic, J.~Er\"{o}, C.~Fabjan, M.~Friedl, R.~Fr\"{u}hwirth, V.M.~Ghete, J.~Hammer\cmsAuthorMark{1}, S.~H\"{a}nsel, M.~Hoch, N.~H\"{o}rmann, J.~Hrubec, M.~Jeitler, W.~Kiesenhofer, M.~Krammer, D.~Liko, I.~Mikulec, M.~Pernicka, H.~Rohringer, R.~Sch\"{o}fbeck, J.~Strauss, A.~Taurok, F.~Teischinger, P.~Wagner, W.~Waltenberger, G.~Walzel, E.~Widl, C.-E.~Wulz
% ...
\vskip\cmsinstskip
% ...
1:~~Also at CERN, European Organization for Nuclear Research, Geneva, Switzerland\\
% ...

\end{sloppypar}
\end{document}